\documentclass[conference,letterpaper]{IEEEtran}

\usepackage[numbers,sort&compress]{natbib}

\makeatletter

\usepackage{etex}

\usepackage{zref-savepos}

\newcounter{mnote}

\def\xmarginnote{%
  \xymarginnote{\hskip -\marginparsep \hskip -\marginparwidth}}

\def\ymarginnote{%
  \xymarginnote{\hskip\columnwidth \hskip\marginparsep}}

\long\def\xymarginnote#1#2{%
\vadjust{#1%
\smash{\hbox{{%
        \hsize\marginparwidth
        \@parboxrestore
        \@marginparreset
\footnotesize #2}}}}}

\def\mnoteson{%
\gdef\mnote##1{\refstepcounter{mnote}\label{##1}%
  \zsavepos{##1}%
  \ifnum20432158>\number\zposx{##1}%
  \xmarginnote{{\color{blue}\bf $\langle$\arabic{mnote}$\rangle$}}%
  \else
  \ymarginnote{{\color{blue}\bf $\langle$\arabic{mnote}$\rangle$}}%
  \fi%
}
  }
\gdef\mnotesoff{\gdef\mnote##1{}}
\mnoteson
\mnotesoff

\newcommand{\figref}[1]{Fig.~\ref{#1}}

\usepackage{framed}
\usepackage{subcaption}
\usepackage{comment}
\usepackage[svgnames,dvipsnames]{xcolor}
\usepackage[all]{xy}
\usepackage{tikz}
\usetikzlibrary{positioning,matrix,through,calc,arrows,fit,shapes,decorations.pathreplacing,decorations.markings,}

\tikzstyle{block} = [draw,fill=blue!20,minimum size=2em]

\usepackage{qsymbols,amssymb,mathrsfs}
\usepackage{amsmath}
\usepackage[standard,thmmarks]{ntheorem}
\theoremstyle{plain}
\theoremsymbol{\ensuremath{_\vartriangleleft}}
\theorembodyfont{\itshape}
\theoremheaderfont{\normalfont\bfseries}
\theoremseparator{}

\theoremstyle{nonumberplain}
\theoremheaderfont{\scshape}
\theorembodyfont{\normalfont}
\theoremsymbol{\ensuremath{_\blacktriangleleft}}

\theoremnumbering{arabic}
\theoremstyle{plain}
\usepackage{latexsym}
\theoremsymbol{\ensuremath{_\Box}}
\theorembodyfont{\itshape}
\theoremheaderfont{\normalfont\bfseries}
\theoremseparator{}

\theorembodyfont{\upshape}
\theoremprework{\bigskip\hrule}
\theorempostwork{\hrule\bigskip}

\usepackage[overload]{empheq} 

\let\iftwocolumn\if@twocolumn
\g@addto@macro\@twocolumntrue{\let\iftwocolumn\if@twocolumn}
\g@addto@macro\@twocolumnfalse{\let\iftwocolumn\if@twocolumn}

\mathtoolsset{showonlyrefs=false,showmanualtags}
\let\underbrace\LaTeXunderbrace 
\let\overbrace\LaTeXoverbrace
\renewcommand{\eqref}[1]{\textup{(\refeq{#1})}} 
\newtagform{brackets}[]{(}{)}   
\usetagform{brackets}

\usepackage[Smaller]{cancel}

\PassOptionsToPackage{breaklinks,letterpaper,hyperindex=true,backref=false,bookmarksnumbered,bookmarksopen,linktocpage,colorlinks,linkcolor=BrickRed,citecolor=OliveGreen,urlcolor=Blue,pdfstartview=FitH}{hyperref}
\usepackage{hyperref}

\usepackage{graphicx,psfrag}
\graphicspath{{figure/}{image/}} 

\usepackage{diagbox} 

\usepackage{algorithm2e}
\usepackage{listings} 
\lstdefinelanguage{Maple}{
  morekeywords={proc,module,end, for,from,to,by,while,in,do,od
    ,if,elif,else,then,fi ,use,try,catch,finally}, sensitive,
  morecomment=[l]\#,
  morestring=[b]",morestring=[b]`}[keywords,comments,strings]
\DeclareMathAlphabet{\mathpzc}{OT1}{pzc}{m}{it}
\usepackage{upgreek} 
\usepackage{dsfont}  

\def\multi@nostar#1#2{%
  \expandafter\def\csname multi#1\endcsname##1{%
    \if ##1.\let\next=\relax \else
    \def\next{\csname multi#1\endcsname}     
    \expandafter\newcommand\csname #1##1\endcsname{#2}
    \fi\next}}

\def\multi@star#1#2{%
  \expandafter\def\csname #1\endcsname##1{#2}
  \multi@nostar{#1}{#2}
}

\newcommand{\multi}{%
  \@ifstar \multi@star \multi@nostar}

\multi*{rm}{\mathrm{#1}}
\multi*{mc}{\mathcal{#1}}
\multi*{op}{\mathop {\operator@font #1}}
\multi*{ds}{\mathds{#1}}
\multi*{set}{\mathcal{#1}}
\multi*{rsfs}{\mathscr{#1}}
\multi*{pz}{\mathpzc{#1}}
\multi*{M}{\boldsymbol{#1}}
\multi*{R}{\mathsf{#1}}
\multi*{RM}{\M{\R{#1}}}
\multi*{bb}{\mathbb{#1}}
\multi*{td}{\tilde{#1}}
\multi*{tR}{\tilde{\mathsf{#1}}}
\multi*{trM}{\tilde{\M{\R{#1}}}}
\multi*{tset}{\tilde{\mathcal{#1}}}
\multi*{tM}{\tilde{\M{#1}}}
\multi*{baM}{\bar{\M{#1}}}
\multi*{ol}{\overline{#1}}

\multirm  ABCDEFGHIJKLMNOPQRSTUVWXYZabcdefghijklmnopqrstuvwxyz.
\multiol  ABCDEFGHIJKLMNOPQRSTUVWXYZabcdefghijklmnopqrstuvwxyz.
\multitR   ABCDEFGHIJKLMNOPQRSTUVWXYZabcdefghijklmnopqrstuvwxyz.
\multitd   ABCDEFGHIJKLMNOPQRSTUVWXYZabcdefghijklmnopqrstuvwxyz.
\multitset ABCDEFGHIJKLMNOPQRSTUVWXYZabcdefghijklmnopqrstuvwxyz.
\multitM   ABCDEFGHIJKLMNOPQRSTUVWXYZabcdefghijklmnopqrstuvwxyz.
\multibaM   ABCDEFGHIJKLMNOPQRSTUVWXYZabcdefghijklmnopqrstuvwxyz.
\multitrM   ABCDEFGHIJKLMNOPQRSTUVWXYZabcdefghijklmnopqrstuvwxyz.
\multimc   ABCDEFGHIJKLMNOPQRSTUVWXYZabcdefghijklmnopqrstuvwxyz.
\multiop   ABCDEFGHIJKLMNOPQRSTUVWXYZabcdefghijklmnopqrstuvwxyz.
\multids   ABCDEFGHIJKLMNOPQRSTUVWXYZabcdefghijklmnopqrstuvwxyz.
\multiset  ABCDEFGHIJKLMNOPQRSTUVWXYZabcdefghijklmnopqrstuvwxyz.
\multirsfs ABCDEFGHIJKLMNOPQRSTUVWXYZabcdefghijklmnopqrstuvwxyz.
\multipz   ABCDEFGHIJKLMNOPQRSTUVWXYZabcdefghijklmnopqrstuvwxyz.
\multiM    ABCDEFGHIJKLMNOPQRSTUVWXYZabcdefghijklmnopqrstuvwxyz.
\multiR    ABCDEFGHIJKL NO QR TUVWXYZabcd fghijklmnopqrstuvwxyz.
\multibb   ABCDEFGHIJKLMNOPQRSTUVWXYZabcdefghijklmnopqrstuvwxyz.
\multiRM   ABCDEFGHIJKLMNOPQRSTUVWXYZabcdefghijklmnopqrstuvwxyz.

\newcommand{\dotleq}{\buildrel \textstyle  .\over {\smash{\lower
      .2ex\hbox{\ensuremath\leqslant}}\vphantom{=}}}
\newcommand{\dotgeq}{\buildrel \textstyle  .\over {\smash{\lower
      .2ex\hbox{\ensuremath\geqslant}}\vphantom{=}}}

\newcommand{\bM}{\begin{bmatrix}}
\newcommand{\eM}{\end{bmatrix}}
\newcommand{\bSM}{\left[\begin{smallmatrix}}
\newcommand{\eSM}{\end{smallmatrix}\right]}
\renewcommand*\env@matrix[1][*\c@MaxMatrixCols c]{%
  \hskip -\arraycolsep
  \let\@ifnextchar\new@ifnextchar
  \array{#1}}

\newqsymbol{`N}{\mathbb{N}}
\newqsymbol{`R}{\mathbb{R}}
\newqsymbol{`P}{\mathbb{P}}
\newqsymbol{`Z}{\mathbb{Z}}

\newqsymbol{`|}{\mid}
\newqsymbol{`8}{\infty}
\newqsymbol{`1}{\left}
\newqsymbol{`2}{\right}
\newqsymbol{`6}{\partial}
\newqsymbol{`0}{\emptyset}
\newqsymbol{`-}{\leftrightarrow}
\newqsymbol{`<}{\langle}
\newqsymbol{`>}{\rangle}

\DeclarePairedDelimiter\abs{\lvert}{\rvert}

\DeclarePairedDelimiter\Set{\{}{\}}
\newcommand{\imod}[1]{\allowbreak\mkern10mu({\operator@font mod}\,\,#1)}

\newcommand{\threecols}[3]{
\hbox to \textwidth{%
      \normalfont\rlap{\parbox[b]{\textwidth}{\raggedright#1\strut}}%
        \hss\parbox[b]{\textwidth}{\centering#2\strut}\hss
        \llap{\parbox[b]{\textwidth}{\raggedleft#3\strut}}%
    }
}

\newcommand{\reason}[2][\relax]{
  \ifthenelse{\equal{#1}{\relax}}{
    \left(\text{#2}\right)
  }{
    \left(\parbox{#1}{\raggedright #2}\right)
  }
}

\newcommand{\utag}[2]{\mathop{#2}\limits^{\text{(#1)}}}
\newcommand{\uref}[1]{(#1)}

\let\SavedDoubleVert\relax
\let\protect\relax
{\catcode`\|=\active
  \xdef\extendvert{\protect\expandafter\noexpand\csname extendvert \endcsname}
  \expandafter\gdef\csname extendvert \endcsname#1{\mskip-5mu \left.%
      \ifx\SavedDoubleVert\relax \let\SavedDoubleVert\|\fi
     \:{\let\|\SetDoubleVert
       \mathcode`\|32768\let|\SetVert
     #1}\:\right.\mskip-5mu}
}
\def\SetVert{\@ifnextchar|{\|\@gobble}
    {\egroup\;\mid@vertical\;\bgroup}}
\def\SetDoubleVert{\egroup\;\mid@dblvertical\;\bgroup}

\begingroup
 \edef\@tempa{\meaning\middle}
 \edef\@tempb{\string\middle}
\expandafter \endgroup \ifx\@tempa\@tempb
 \def\mid@vertical{\middle|}
 \def\mid@dblvertical{\middle\SavedDoubleVert}
\else
 \def\mid@vertical{\mskip1mu\vrule\mskip1mu}
 \def\mid@dblvertical{\mskip1mu\vrule\mskip2.5mu\vrule\mskip1mu}
\fi

\makeatother

\usepackage{fouridx}
\usepackage{framed}
\usetikzlibrary{positioning,matrix}

\usepackage{paralist}
\usepackage{enumerate}

\usepackage[normalem]{ulem}

\numberwithin{equation}{section}
\makeatletter
\@addtoreset{equation}{section}
\renewcommand{\theequation}{\arabic{section}.\arabic{equation}}
\@addtoreset{Theorem}{section}
\renewcommand{\theTheorem}{\arabic{section}.\arabic{Theorem}}
\@addtoreset{Lemma}{section}
\renewcommand{\theLemma}{\arabic{section}.\arabic{Lemma}}
\@addtoreset{Corollary}{section}
\renewcommand{\theCorollary}{\arabic{section}.\arabic{Corollary}}
\@addtoreset{Example}{section}
\renewcommand{\theExample}{\arabic{section}.\arabic{Example}}
\@addtoreset{Remark}{section}
\renewcommand{\theRemark}{\arabic{section}.\arabic{Remark}}
\@addtoreset{Proposition}{section}
\renewcommand{\theProposition}{\arabic{section}.\arabic{Proposition}}
\@addtoreset{Definition}{section}
\renewcommand{\theDefinition}{\arabic{section}.\arabic{Definition}}
\@addtoreset{Claim}{section}

\@addtoreset{Subclaim}{Theorem}
\renewcommand{\theSubclaim}{\theTheorem\Alph{Subclaim}}
\makeatother

\newenvironment{lbox}{
  \setlength{\FrameSep}{1.5mm}
  \setlength{\FrameRule}{0mm}
  \MakeFramed {\FrameRestore}}%
{\endMakeFramed}

\newenvironment{ybox}{
	\setlength{\FrameSep}{1.5mm}
	\setlength{\FrameRule}{0mm}
  \MakeFramed {\FrameRestore}}%
{\endMakeFramed}

\newenvironment{gbox}{
	\setlength{\FrameSep}{1.5mm}
\setlength{\FrameRule}{0mm}
  \MakeFramed {\FrameRestore}}%
{\endMakeFramed}

{\endMakeFramed}

 {\endMakeFramed}

\usepackage{enumitem}

\usepackage{etoolbox}

\let\theparentequation\theequation
\patchcmd{\theparentequation}{equation}{parentequation}{}{}

\renewenvironment{subequations}[1][]{
	\refstepcounter{equation}%
	\setcounter{parentequation}{\value{equation}}
	\setcounter{equation}{0}
	\def\theequation{\theparentequation\alph{equation}}%
	\let\parentlabel\label
	\ifx\\#1\\\relax\else\label{#1}\fi
	\ignorespaces
}{%
	\setcounter{equation}{\value{parentequation}}
	\ignorespacesafterend
}

\newcommand*{\nextParentEquation}[1][]{
	\refstepcounter{parentequation}
	\setcounter{equation}{0}
	\ifx\\#1\\\relax\else\parentlabel{#1}\fi
}

\newcommand{\RCO}{R_{\op{CO}}}
\newcommand{\RS}{R_{\op{S}}}
\newcommand{\rK}{r_{\op{K}}}
\newcommand{\CS}{C_{\op{S}}}

\usepackage[outline]{contour}
\contourlength{1.2pt}

\title{Secret Key Agreement under\\ Discussion Rate Constraints}

\author{Chung Chan, Manuj Mukherjee, Navin Kashyap and Qiaoqiao Zhou
		\thanks{C.\ Chan (email: cchan@inc.cuhk.edu.hk),
			 and Q.\ Zhou are with the Institute of Network Coding at the
			Chinese University of Hong Kong, the Shenzhen Key Laboratory of
			Network Coding Key Technology and Application, China, and the
			Shenzhen Research Institute of the Chinese University of Hong
			Kong.
		}
		\thanks{N.\ Kashyap and M.\ Mukherjee are with the Department of Electrical Communication Engineering at the Indian Institute of Science, Bangalore.}
		\thanks{The work described in this paper was supported by a grant from University Grants Committee of the Hong Kong Special Administrative Region, China (Project No. AoE/E-02/08), and supported partially by a grant from Shenzhen Science and Technology Innovation Committee (JSGG20160301170514984), the Chinese University of Hong Kong (Shenzhen), China.}
		\thanks{The work of C.\ Chan was supported in part by The Vice-Chancellor's One-off Discretionary Fund of The Chinese University of Hong Kong (Project Nos. VCF2014030 and VCF2015007), and a grant from the University
			Grants Committee of the Hong Kong Special Administrative Region,
			China (Project No. 14200714).}
		\thanks{The work of N.\ Kashyap and M.\ Mukherjee is supported in part by a Swarnajayanti Fellowship awarded to N. Kashyap by the Department of Science \& Technology (DST), Government of India.}
	}

\begin{document}

\IEEEoverridecommandlockouts
\maketitle

\begin{abstract}
For the multiterminal secret key agreement problem, new single-letter lower bounds are obtained on the public discussion rate required to achieve any given secret key rate below the secrecy capacity. The results apply to general source model without helpers or wiretapper's side information but can be strengthened for hypergraphical sources. In particular, for the pairwise independent network, the results give rise to a complete characterization of the maximum secret key rate achievable under a constraint on the total discussion rate.
\end{abstract} 



\section{Introduction}
\label{sec:introduction}

We consider the multiterminal secret key agreement by public discussion in \cite{csiszar04} under the source model without helpers or wiretapper's side information. While the maximum achievable secret key rate with unlimited public discussion, called the secrecy capacity, was characterized in \cite{csiszar04} using an achieving scheme through the omniscience of the source, it was pointed out~\cite{csiszar04} that the proposed scheme may not achieve the minimum public discussion rate, referred to as the communication complexity. While a multi-letter characterization was derived in \cite{tyagi13} for the $2$-user case, a computable single-letter characterization is a challenging open problem.

Simpler versions of the problem have been considered, such as the introduction of the vocality constraints in \cite{amin10a,mukherjee14,zhang15}. Using the result of \cite{amin10a} with silent users and viewing the secrecy capacity as the multivariate mutual information measure (MMI)~\cite{chan15mi}, these simpler problems can be resolved completely~\cite{chan16so}. Combining the idea of Wyner common information and the MMI, a multi-letter lower bound on the communication complexity was derived in \cite{MKS16}. For the pairwise independent network (PIN)~\cite{nitinawarat10}, the bound leads to a precise single-letter condition in~\cite{MKS16} under which the omniscience strategy in \cite{csiszar04} achieves the communication complexity. The lower bound was further single-letterized and simplified to an easily computable bound in \cite{chan16itw}, where the condition for the optimality of omniscience was also generalized from PINs to hypergraphical sources~\cite{chan10md}, using the idea of decremental secret key agreement in~\cite{chan16isit} for the upper bound~\cite{mukherjee16}. Unfortunately, the lower bound can be loose even for simple PINs. It was also conjectured that the lower bound failed to give the condition for the optimality of omniscience for general sources. 

By resolving the conjecture in \cite{chan16itw}, we discovered new techniques that can improve the lower bound further. Although the techniques are also based on the idea of MMI, they work quite differently compared to the idea of Wyner common information~\cite{MKS16}. 
We apply these techniques to obtain an outer bound on the region of achievable secret key rate and discussion rate tuples. In particular, for PIN models on trees our outer bound turns out to be an exact characterization. In contrast with the rate region characterized in \cite{LCV16} for two terminals using the idea of two-way interactive source coding \cite{Kaspi}, the result is the first instance of an exact and easily computable characterization for the case with at least three terminals with unlimited number of rounds of interactive discussion. We also use the outer bound to characterize
the communication complexity, and more generally, the maximum secret key rate achievable under any given total discussion rate, referred to as the rate-constrained secrecy capacity.

\section{Motivation}
\label{sec:motivation}
\begin{figure}
	\centering
	\tikzstyle{dot}=[circle,draw=gray!80,fill=gray!20,thick,inner
	sep=2pt,minimum size=11pt]
	\begin{tikzpicture}
	\node[dot,label={[label distance=0em]above:{$\RZ_1$}}] (s) at(-2,0) {$1$};
	\node[dot,label={[label distance=0em]above:{$\RZ_2$}}] (a) at(0,0) {$2$} ;
	\node[dot,label={[label distance=0em]above:{$\RZ_3$}}] (t) at(2,0) {$3$};
	\draw[] (-1.77,0) -- (-.25,0);
	\node [] at (-1,0.20) {\scriptsize $\RX_{\rm{a}}$};     
	\draw[] (0.25,0.1) -- (1.77,0.1);
	\node [] at (1,0.30) {\scriptsize $\RX_{\rm{b}}$};     
	\draw[] (0.25,-0.1) -- (1.77,-0.1);
	\node [] at (1,-0.30) {\scriptsize $\RX_{\rm{c}}$};     
	\end{tikzpicture}
	\caption{The graphical representation of the PIN~\eqref{eq:mot:src}. Each edge corresponds to an independent random variable observed by the incident nodes.}
	\label{fig:mot}
\end{figure}
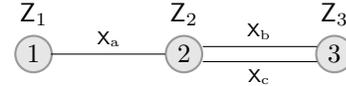
We first motivate the idea of secret key agreement and the main results informally using a simple example. Let $\RX_{\rm{a}}$, $\RX_{\rm{b}}$ and $\RX_{\rm{c}}$ be uniformly random  and independent bits, and define
\begin{equation}
\label{eq:mot:src}
\begin{aligned}
\RZ_1 & :=\RX_{\rm{a}}\\
\RZ_2 & :=(\RX_{\rm{a}},\RX_{\rm{b}},\RX_{\rm{c}})\\
\RZ_3 & :=(\phantom{\RX_{\rm{a}},}\:\RX_{\rm{b}},\RX_{\rm{c}}).
\end{aligned}
\end{equation}
Consider 3 users 1, 2 and 3 observing $\RZ_1$, $\RZ_2$ and $\RZ_3$ respectively in private. The private source $(\RZ_1,\RZ_2,\RZ_3)$ is called a PIN \cite{nitinawarat-ye10,nitinawarat10} in the sense that its statistical dependency can be described by a (multi-)graph as shown in \figref{fig:mot} with the nodes representing the users, $\RX_{\rm{a}}$ represented by an edge incident on nodes~1 and 2, and $\RX_{\rm{b}}$ and $\RX_{\rm{c}}$ represented by two edges incident on nodes 2 and 3.

If user 2 reveals $\RF:=\RX_{\rm{a}}\oplus\RX_{\rm{b}}$ in public so that everyone can observe it, then user 3 can recover $\RX_{\rm{a}}$ as $\RF\oplus\RX_{\rm{b}}$. $\RK:=\RX_{\rm{a}}$ is called a secret key bit generated by the public discussion $\RF$ because $\RK$ is not only recoverable by all users but also uniformly random and independent of the public discussion $\RF$. A general asymptotic secret key agreement protocol by interactive public discussion was formulated in \cite{csiszar04}, where the maximum achievable key rate, called the \emph{secrecy capacity} and denoted by $\CS$, was characterized by a single-letter linear program. For the current example, it is easy to see that $\CS=1$, since user 1 observes at most 1~bit in private and $1$~bit of secret key is achievable by the above discussion scheme. 

A quantity of interest but not characterized in \cite{csiszar04} is the smallest public discussion rate required to achieve the secrecy capacity, called the \emph{communication complexity} and denoted by $\RS$. For the current example, $\RS\leq 1$ because the above capacity-achieving discussion $\RF$ is 1~bit. However, the precise characterization of $\RS$ has been unknown even for the current simple example.

In this work, we introduce new techniques that not only implies $\RS=1$ for the current example but also characterizes the maximum key rate under a total public discussion rate $R\geq0$, called the rate-constrained secrecy capacity and denoted by $\CS(R)$. For the current example, it will follow that 
\begin{align}
\CS(R)=\min\{R,1\}.\label{eq:CSR:mot}
\end{align}
Although it is easy to see that $\CS(0)\geq 0$ and $\CS(R)=1$, for $R\geq 1$, and that $\CS(R)\geq\min\{R,1\}$ by time sharing, proving the reverse inequality is non-trivial and calls for new techniques not covered by \cite{MKS16,chan16itw}. Indeed, our techniques will also imply that only user 2 needs to discuss in public, and so a secret key rate of $\rK\in [0,1]$ is achievable by a discussion rate tuple $(r_1,r_2,r_3)$ iff they belong to the region
\begin{align}
	\begin{split}
	\rsfsR&=\Set{(\rK,(r_1,r_2,r_3))\mid \rK\in [0,1],\\
		&\kern8em r_1\geq 0, r_2\geq \rK, r_3\geq 0}.\end{split}\label{eq:R:mot}
\end{align}
This matches our intuition, since users 1 and 3 have independent private observations, i.e., $\RZ_1$ is independent of $\RZ_3$, and so only user 2 can help them share a non-trivial secret key. It turns out that the techniques apply to more general source model with private randomization and interactive discussion allowed as in \cite{csiszar04}. It also completely characterizes $\CS(R)$ for the PIN model.

\section{Problem formulation}
\label{sec:problem}

We consider the multiterminal secret key agreement \cite{csiszar04} without helpers or wiretapper's side information. It involves a finite set 
$V:=[m]:=\{1,2,\ldots,m\}$
of $m\geq 2$ users. The users have access to a private (discrete memoryless multiple) source denoted by the random vector
\begin{align*}
	\RZ_V & := (\RZ_i|i\in V)\sim P_{\RZ_V} \text{ taking values from }\\
	Z_V & := \prod\nolimits_{i\in V} Z_i, \text{ assumed to be finite.}
\end{align*}
N.b., capital letters in sans serif font are used for random variables and the corresponding capital letters in the usual math italic font denote the alphabet sets. $P_{\RZ_V}$ denotes the joint distribution of $\RZ_i$'s. The protocol can be divided into the following phases:
\begin{lbox}
\noindent\underline{Private observation:} Each user $i\in V$ observes an $n$-sequence 
$$
\RZ_i^n:=(\RZ_{it}|t\in[n])=(\RZ_{i1},\RZ_{i2},\ldots,\RZ_{in})
$$
i.i.d. generated from the source $\RZ_i$ for some block length $n$.

\noindent\underline{Private randomization:} Each user $i\in V$ generates a random variable $\RU_i$ independent of the private source, i.e.,
\begin{align}
	\label{eq:U}
H(\RU_V|\RZ_V)=\sum_{i\in V}H(\RU_i).
\end{align}
For convenience, we denote the entire private observation of user $i\in V$ as 
\begin{align}
\tRZ_i:=(\RU_i,\RZ_i^n).\label{eq:tRZ}
\end{align}

\noindent\underline{Public discussion:}  Using a public authenticated noiseless channel, each user $i\in V$ broadcasts a message in round $t$
\begin{subequations}
	\label{eq:discussion}
\begin{alignat}{2}
\RF_{it}&:=f_{it}(\tRZ_i,\tRF_{it})&\kern1em&\text {where} \label{eq:Fit}\\
\tRF_{it} &:=(\RF_{[i-1]t},\RF_V^{t-1}),\label{eq:tFit}
\end{alignat}
$t\in[\ell]$ for some positive integer $\ell$ number of rounds, $\RF_{[i-1]t}$ consists of the previous messages broadcast in the same round, while $\RF_V^{t-1}$ denotes the messages broadcast in the previous rounds. Without loss of generality, we assume this interactive discussion is conducted in the ascending order of user indices. We also write
\begin{align}
	\RF_i & := \RF_{i[\ell]}=(\RF_{it}|t\in[\ell])\label{eq:Fi}\\
	\RF & := \RF_V = (\RF_i|i\in V)\label{eq:F}
\end{align}
to denote the aggregate message from user $i\in V$ and the aggregation of the messages from all users respectively.
\end{subequations}
\noindent\underline{Key generation:} A random variable $\RK$, called the secret key, is required to satisfy the recoverability constraint that
\begin{equation}
\lim_{n\to\infty}\text{Pr}(\exists i\in V, \RK\neq\theta_i(\tRZ_i,\RF))=0, \label{eq:recover}
\end{equation}
for some function $\theta_i$, and the secrecy constraint that
\begin{equation}
\lim_{n\to\infty}\frac{1}{n}`1[\log\abs{K}-H(\RK|\RF)`2]=0, \label{eq:secrecy}
\end{equation}
where $K$ denotes the finite alphabet set of possible key values.
\end{lbox}

\begin{Definition}\label{def:CS(R)}
	Given the private source $\RZ_V$, a secret key rate $\rK$ is achievable by the public discussion rate tuple $r_V:=(r_i|i\in V)$ iff
	\begin{equation}
	\rK\leq\liminf_{n\to\infty}\frac{1}{n}\log\abs{K} \text{ and } r_i\geq\limsup_{n\to\infty}\frac{1}{n}\log\abs{F_i}, \label{eq:rate}
	\end{equation}
	in addition to \eqref{eq:recover} and \eqref{eq:secrecy}. The set of achievable $(\rK,r_V)$ is denoted by $\rsfsR$.
	The \emph{rate-constrained secrecy capacity} is defined for $R\geq 0$ as 
	\begin{equation}
	\CS(R):=\max\{\rK\mid (\rK,r_V)\in\rsfsR, r(V)\leq R\},\label{CS(R)}
	\end{equation}
	where, for convenience, $r(B):=\sum_{i\in B} r_i$ for $B\subseteq V$.
\end{Definition}

\begin{Proposition}\label{prop:CS(R)}
	$\CS(R)$ is continuous, non-decreasing and concave for $R\geq 0$.
\end{Proposition}

\begin{Proof}
	Continuity is because the liminf and limsup in \eqref{eq:rate} always exist, since $\CS(R)$ is bounded within $[0,H(\RZ_V)]$. The monotonicity is obvious, and concavity follows from the usual time sharing argument.
\end{Proof}

The \emph{unconstrained secrecy capacity} defined and characterized in \cite{csiszar04} is the special case
\begin{align}
	\CS&:=\lim_{R\to\infty}\CS(R) \label{eq:CS}\\
	&= \CS(\RCO) = H(\RZ_V)-\RCO\notag 
\end{align}
where $\RCO$ is the \emph{smallest rate of communication for omniscience}, characterzied in \cite{csiszar04} by the linear program
\begin{align}
	\kern-.5em \RCO=\min\{r(V)\mid r(B)\geq H(\RZ_B|\RZ_{V`/B}),\forall B\subsetneq V\}.\kern-.5em 
	\label{eq:RCO}
\end{align}
It was also mentioned in \cite{csiszar04} that the unconstrained capacity can be attained by a possibly smaller discussion rate, referred to as the communication complexity
\begin{align}
	\RS & := \min\{r(V)\mid (\CS,r_V)\in\rsfsR\} \label{eq:RS}\\
	& = \min\{R\geq 0\mid \CS(R)=\CS\}\leq\RCO.\notag
\end{align}
Our goal is to characterize or bound $\CS(R)$ and $\rsfsR$ using only single-letter expressions. We will also specialize and strengthen the results to the hypergraphical source model:
\begin{Definition}[Definition~2.4 of \cite{chan10md}]\label{def:hyp}
	$\RZ_V$ is a \emph{hypergraphical source} w.r.t.\  a hypergraph $(V,E,`x)$ with edge functions $`x: E\to2^V`/\{\emptyset\}$ iff, for some independent (hyper)edge variables $\RX_e$ for $e\in E$ with $H(\RX_e)>0$,
	\begin{equation}
	\RZ_i:=(\RX_e\mid  e\in E, i\in`x(e)), \text{ for } i\in V. \label{eq:Xe}
	\end{equation}
	The \emph{weight function} $c:2^V`/\{\emptyset\}\to\mathbb{R}$ of a hypergraphical source is defined as
	\begin{subequations}
		\label{eq:c}
		\begin{align}
		c(B)&:=H(\RX_e\mid e\in E,`x(e)=B) \kern.5em \text{with support}\kern-.5em\\
	\kern-.5em \op{supp}(c)&:=\Set*{B\in 2^V`/\{\emptyset\} \mid c(B)>0}
	\end{align}
	\end{subequations}
\end{Definition}
The PIN model~\cite{nitinawarat10} such as \eqref{eq:mot:src} is an example, where the corresponding hypergraph is the graph in \figref{fig:mot} with weight $c(\Set{1,2})=H(\RX_{\rm{a}})=1$, $c(\Set{2,3})=H(\RX_{\rm{b}},\RX_{\rm{c}})=2$ and $0$ otherwise.
\begin{Definition}[\cite{nitinawarat10}]
	$\RZ_V$ is a PIN iff it is hypergraphical w.r.t.\  a graph $(V,E,`x)$ with edge function $`x: E\to V^2`/\{(i,i)\mid i\in V\}$ (i.e., no self loops).
\end{Definition}
For this special source model, there is a protocol in \cite[Proof of Theorem~3.3]{nitinawarat-ye10} that achieves the unconstrained secrecy capacity~\cite[(15),(17)]{nitinawarat-ye10}.
\begin{Proposition}[\mbox{\cite{nitinawarat-ye10,nitinawarat10}}]\label{pro:tree-packing}
	For a PIN with weight $c$, there is a secret key agreement scheme, called the \emph{tree-packing protocol}, which achieves $(\rK,r_V) \in \rsfsR$ with
	\begin{subequations}
	\begin{align}
	\kern-.5em \rK:=\kern-.2em \sum_{j\in [k]} `h_j \kern.5em \text{and} \kern.5em
	r_i:= \kern-.2em \sum_{j\in [k]} (d_{T_j}(i)\kern-.2em -\kern-.2em 1) `h_j \kern.5em \text{for $i\in V$,}\kern-.5em 
	\label{eq:tree-packing:rate}
\end{align}
	where $k$ is a non-negative integer; $`h_j\in `R_+$ is a non-negative real number; $T_j:=(V,\mcE_j)$ is a spanning tree with edge set $\mcE_j\subseteq V^2`/\Set{(i,i)\mid i\in V}$ satisfying
	\begin{align}
		\sum_{j\in [k]:B\in \mcE_j} `h_j \leq c(B)
		\kern1em \forall B\in 2^V`/\Set{`0},
		\label{eq:tree-packing:cons}
	\end{align}
	\end{subequations}
	which is the constraint for fractional tree-packing~\cite{schrijver02}; and $d_{T_j}(i)$ is the degree of node $i$ in $T_j$. Furthermore, the unconstrained secrecy capacity $\CS$ is the maximum $\rK$ over the fractional tree packing $\Set{(`h_j,T_j)\mid i\in [k]}$.
\end{Proposition}
However, it was left as an open problem in \cite{nitinawarat10} whether the above scheme achieves $\RS$. We resolve this in the affirmative by providing a matching converse.

\section{Main results}
\label{sec:results}

We will make use of the following alternative characterization of the unconstrainted secrecy capacity in \cite{chan10md}: For the no-helper case, $\CS=I(\RZ_V)$ where $I(\RZ_V)$ is called the multivariate mutual information (MMI) defined as
\begin{subequations}
	\label{eq:mmi}
\begin{align}
	I(\RZ_V) & := \min_{\mcP\in\Pi'(V)}I_{\mcP}(\RZ_V), \text{ with }\label{eq:I}\\
	I_{\mcP}(\RZ_V) & :=\frac{1}{|\mcP|-1}\biggl[\underbrace{\sum\nolimits_{C\in\mcP}H(\RZ_C)-H(\RZ_V)}_{=D(P_{\RZ_V}\|\prod_{C\in \mcP} P_{\RZ_C})}\biggr]\label{eq:IP}
\end{align}
\end{subequations}
and $\Pi'(V)$ being the set of partitions of $V$ into at least 2 non-empty disjoint subsets of $V$. The conditional versions $I(\RZ_V|\RW')$ and $I_{\mcP}(\RZ_V|\RW')$ are defined in the same way but with the entropy terms conditioned on $\RW'$ in addition. $D(\cdot\|\cdot)$ is the Kullback--Leibler divergence, which is non-negative, and so are $I$ and $I_{\mcP}$. It was pointed out in \cite{chan15mi} that the set of optimal solutions form a lattice w.r.t.\  the partial order $\mcP'\succeq \mcP$ iff
\begin{align*}
	\forall C\in \mcP, \exists C'\in \mcP': C\subseteq C'.
\end{align*}
Hence, there exists a unique finest optimal partition, denoted by $\mcP^*(\RZ_V)$ and referred to as the fundamental partition. Furthermore, both the MMI and the optimal partitions can be computed in strongly polynomial time w.r.t.\ the number of evaluation of the entropies.

In the bivariate case when $V=\Set{1,2}$, the MMI reduces to Shannon's mutual information
\begin{align*}
	I(\RZ_{\Set{1,2}})=I(\RZ_1\wedge \RZ_2)= H(\RZ_1)+H(\RZ_2)-H(\RZ_1,\RZ_2),
\end{align*}
because $\Set{\Set{1},\Set{2}}$ is the unique partition in $\Pi'(\Set{1,2})$ (and is therefore the fundamental partition $\mcP^*(\RZ_{\Set{1,2}})$). 


We begin with some general lower bounds on the public discussion rates:
\begin{Theorem}
	\label{thm:LB}
	For any $(\rK,r_V)\in\rsfsR$, we have 
	\begin{align}
		\label{eq:LB}
		r(V`/B)\geq (\abs{\mcP}-1)[\rK-I_{\mcP}(\RZ_B)]  
	\end{align}
	for any $B\subseteq V$ with size $\abs{B}>1$ and $\mcP\in\Pi'(B)$.
\end{Theorem}

\begin{Proof}
	See Appendix~\ref{sec:proof}.
\end{Proof}

\eqref{eq:LB} is a lower bound on the total discussion rate $r(V`/B)$ of the subset $V`/B$ of users required to achieve a secret key rate of $\rK$, for any choice of subset $B$ of more than one user. Choosing $\mcP$ to be the fundamental partition $\mcP^*(\RZ_B)$ in \eqref{eq:LB}, $I_{\mcP}(\RZ_B)=I(\RZ_B)$, which gives the following lower bound in terms of the MMI.

\begin{Corollary}
	For any $(\rK,r_V)\in\rsfsR$, we have 
	\begin{align}
		\label{eq:MMI:LB}
		r(V`/B)\geq (\abs{\mcP^*(\RZ_B)}-1)[\rK-I(\RZ_B)]  
	\end{align}
	for any $B\subseteq V$ with size $\abs{B}>1$.
\end{Corollary}
Note that $I(\RZ_B)$ in \eqref{eq:MMI:LB} is the secrecy capacity when users in $V`/B$ are removed. Hence, to achieve a secret key rate beyond $I(\RZ_B)$, users in $V`/B$ must discuss. \eqref{eq:MMI:LB} states that the total discussion rate of users in $V`/B$ is at least the additional secret key rate $\rK-I(\RZ_B)$ amplified by a factor of $\abs{\mcP^*(\RZ_B)}-1\geq 1$.

Applying~\eqref{eq:LB} to the example in Section~\ref{sec:motivation} with $B=\Set{1,3}, \mcP=\Set{\Set{1},\Set{3}}$ (or simply \eqref{eq:MMI:LB}), we have 
\begin{align}
	r_2 \geq (2-1)[\rK-I(\RZ_1 \wedge \RZ_3)]=\rK \label{eq:LB:mot}
\end{align}
This is achievable as mentioned in Section~\ref{sec:motivation} by time sharing between $(\rK,(r_1,r_2,r_3))=(0,(0,0,0))$ and $(1,(0,1,0))\in \rsfsR$. Since $\CS=I(\RZ_{\Set{1,2,3}})\leq I(\RZ_{\Set{1,2}}\wedge\RZ_3)=1$ and is achievable, we have \eqref{eq:R:mot} as the achievable rate region $\rsfsR$. More generally, 
\begin{Theorem}
	\label{thm:PIN:T}
	For PIN with weight $c$ such that $\op{supp}(c)$, defined in~\eqref{eq:c}, forms a spanning tree, we have
	\begin{subequations}
		\begin{align}
			\begin{split}
			\rsfsR&=\{(\rK,r_V)\mid \rK\in[0,\CS],\\
			&\kern4em r_i \geq `1(d(i)-1`2)\rK, i\in V \}, \kern1em \text{where} 
			\end{split}\label{eq:PIN:T:R}\\
			\CS&=\min`1\{c(\Set{i,j})\mid \Set{i,j}\in \op{supp}(c)`2\},\label{eq:PIN:T:CS}
		\end{align}
		and $d(i)$ is the degree of node $i$ in the spanning tree.
	\end{subequations}
\end{Theorem}

\begin{Proof}
	Since the source model forms a Markov tree w.r.t.\  the spannnig tree given by $\op{supp}(c)$, the unconstrained secrecy capacity~\eqref{eq:PIN:T:CS} follows from \cite[(36)]{csiszar04}. 
	
	To prove \eqref{eq:PIN:T:R},
	consider any PIN with weight function $c$ such that $\op{supp}(c)$ forms a spanning tree. For any $i\in V$, choose $B=V`/\Set{i}$ and let $\mcP$ be the connected components of the spanning tree after node $i$ and its incident edges are removed. It follows that $\mcP\in \Pi'(B)$ with
	\begin{align*}
		\abs{\mcP}&=d(i)\kern1em \text{ and }\kern1em 
		I_{\mcP}(\RZ_{B})=0
	\end{align*}
	due to the fact that $\op{supp}(c)$ forms a spanning tree. By~\eqref{thm:LB} in Theorem~\ref{eq:LB}, we have
	\begin{align*}
		r_i&\geq (\abs{\mcP}-1)`1[\rK-I_{\mcP}(\RZ_{B})`2]\\
		&= `1(d(i)-1`2)\rK.
	\end{align*}
	The lower bound is achievable by Proposition~\ref{pro:tree-packing}, hence completing the proof of \eqref{eq:PIN:T:R}.
\end{Proof}

The current example has a weight function $c$ with $$\op{supp}(c)=\Set{\Set{1,2},\Set{2,3}},$$ which is a spanning tree with node degrees given by
\[d(1)=d(3)=1\kern1em \text {and}\kern1em d(2)=2, \]
which gives the lower bound~\eqref{eq:LB:mot} and hence the region in~\eqref{eq:R:mot}. The capacity is the minimum edge weight, i.e.,
$$\CS=\min`1\{c(\Set{1,2}), c(\Set{2,3})`2\}=\min\Set{1,2}=1.$$

\begin{figure}
\centering
\tikzstyle{dot}=[circle,draw=gray!80,fill=gray!20,thick,inner
sep=2pt,minimum size=11pt]
\tikzstyle{point}=[draw,circle,minimum size=.2em,inner sep=0, outer sep=.2em]
\tikzstyle{+}=[draw,fill=white,circle,minimum size=.8em,inner sep=0pt]
\begin{tikzpicture}[x=.6em,y=.6em,>=latex]
\foreach \x/\angle/\lb in {3/210/{$\RZ_\x$},1/90/{$\RZ_\x$},2/-30/{$\RZ_\x$}}
{
	\path (\angle:5) node (\x) [dot,label={[label distance=0em]\angle:{\lb}}] {$\x$};
}
\foreach \x/\y/\lp/\lb in {1/2/right/$\RX_{\rm{a}}$,
	2/3/below/$\RX_{\rm{b}}$,
	1/3/left/$\RX_{\rm{c}}$}
\draw[-] (\x) to node [label=\lp:{\scriptsize\lb}]{} (\y);
\end{tikzpicture}
\caption{The triangle PIN defined in~\eqref{eq:triangle}.}
\label{fig:triangle}
\end{figure}
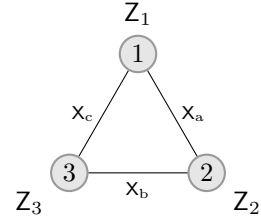

Unfortunately, the lower bound~\eqref{eq:LB} can be loose for PIN with cycles. E.g., consider a triangle PIN with
 $V:=[3]$ and
\begin{equation}
	\label{eq:triangle}
	\begin{aligned}
		\RZ_1&:=(\RX_{\rm{a}},\kern1.6em\RX_{\rm{c}})\\
		\RZ_2&:=(\RX_{\rm{a}},\RX_{\rm{b}}\kern1.7em)\\
		\RZ_3&:=(\kern1.7em\RX_{\rm{b}},\RX_{\rm{c}})
	\end{aligned}
\end{equation}
where $\RX_{\rm{a}},\RX_{\rm{b}},\RX_{\rm{c}}$ are independent uniformly random bits. This a PIN with correlation represented by a triangle in \figref{fig:triangle}. It follows from \eqref{eq:CS}, \eqref{eq:RCO} and \eqref{eq:RS} that 
$$\CS=\RCO=1.5\geq \RS.$$
In particular, the secret key rate of $1$ is achievable by the scheme described in Section~\ref{sec:motivation}.

Applying \eqref{eq:LB} with $B=\Set{1,3}$ and $\mcP=\Set{\Set{1},\Set{3}}$ as before, 
\begin{align*}
	r_2\geq \rK-I(\RZ_1\wedge\RZ_3)=\rK-1.
\end{align*}
This is the best possible bound involving $r_2$ over all possible choices of $B$ and $\mcP$, but it is trivial when $\rK\leq 1$. By symmetry, the best bounds for $r_1$ and $r_3$ are also trivial when $\rK\leq 1$. 

Nevertheless, we discovered a different bounding technique that can give a non-trivial bound in the above case, by exploiting the hypergraphical dependency structure of the source: 
\begin{Theorem}
	\label{thm:LB:hyp}
	For hypergraphical source, we have $(\rK,r_V)\in\rsfsR$ only if 
	\begin{subequations}
		\label{eq:LB:hyp}
		\begin{align}
			&\alpha(\mcP) r(V)\geq `1[1-\alpha(\mcP)`2]\rK  \kern1em \forall \mcP\in \Pi'(V),\kern1em\text {where}\label{eq:LB:hyp:R}\\
			&\alpha(\mcP):=\frac{\max_{e\in E}\abs*{\Set{C\in\mcP\mid C\cap`x(e)\neq `0} }-1}{\abs{\mcP}-1}\label{eq:LB:hyp:alpha}
		\end{align} 
	and $`x$ is the edge function of the hypergraph in~\eqref{eq:Xe}. 
	\end{subequations}
\end{Theorem}
N.b., it is easy to see that $`a(\mcP)\in [0,1]$ because the maximization in the numerator of \eqref{eq:LB:hyp:alpha} is the maximum number of blocks in $\mcP$ that an edge $e\in E$ can intersect, which is between $1$ and $\abs {\mcP}$. If $`a(\mcP)=0$ for some $\mcP\in \Pi'(V)$, then \eqref{eq:LB:hyp:R} becomes $r_K\leq 0$, i.e., $\CS=0$. This happens when no edge crosses $\mcP$, i.e., the source corresponds to a disconnected hypergraph. 
\begin{Proof}
	See Appendix~\ref{sec:proof:hyp}.
\end{Proof}

For the current example, choose $\mcP=\Set{\Set{1},\Set{2},\Set{3}}$. For each edge $e$, $\Big|\Set{C\in\mcP| C\cap`x(e)\neq `0}\Big|$ simplifies to the number of incident nodes, which is always $2$ for graphs. Hence, 
\[ \alpha(\mcP)=\frac{2-1}{3-1}=\frac12 \kern1em \text {and so}\kern1em r(V)\geq \frac{1-\frac12}{\frac12}\rK=\rK.\]
Since $\CS=\RCO=1.5$, the lower bound above is achievable by time-sharing, which gives
\begin{align*}
	\CS(R)=\min\Set{R,1.5} \kern1em\text{and so}\kern1em \RS=1.5.
\end{align*}
Surprisingly, the argument can be extended to any PIN for a complete characterization of the communication complexity as well as the rate-constrained secrecy capacity.

\begin{Theorem}
	For PIN,
	\begin{align}
		\CS(R)=\min`1\{\frac{R}{\abs{V}-2}, \CS`2\},
	\end{align}
	which gives
	$\RS=(\abs{V}-2)\CS$.
\end{Theorem}

\begin{Proof}
	The converse follows from~\eqref{eq:LB:hyp:R} with $\mcP=\Set{\Set{i}| i\in V}$. More precisely, the minimization in the numerator of $`a(\mcP)$ is always equal to $2$ as it is the number of incident nodes of an edge. Hence,
	\begin{align*}
		\alpha(\mcP)&=\frac{1}{\abs{V}-1}\kern1em \text{and so}\\
		r(V)&\geq (\abs{V}-2)\rK\kern1em \text{by \eqref{eq:LB:hyp:R}}.
	\end{align*}
	The lower bound can be shown to be achievable by Proposition~\ref{pro:tree-packing}. With $(\rK,r_V)$ defined in \eqref{eq:tree-packing:rate},
	\begin{align*}
		r(V)&=\sum_{i\in V}\sum_{j=1}^{k}[d_{T_j}(i)-1]\eta_j\\
		&=\sum_{j=1}^{k} `h_j \sum_{i\in V}[d_{T_j}(i)-1] = (\abs{V}-2)\rK,
	\end{align*}
	where the last equality follows from the fact that $\sum_{i\in V} d_{T_j}(i)=\abs {\mcE_j}=\abs {V}-1$ as $T_j$ is a spanning tree.
\end{Proof}

\section{Extensions and challenges}
\label{sec:extension}

While the lower bound~\eqref{eq:LB} can be loose in the presence of cycles, it can be shown to be tight for hypergraphical sources that correspond to hypergraphs that are minimally connected in the sense that removing any edge disconnects the hypergraphs. This generalizes the result of Theorem~\ref{thm:PIN:T} from PINs to hypergraphical sources. Both lower bounds~\eqref{eq:LB} and \eqref{eq:LB:hyp} can also be extended to include helpers. However, it is unclear how one can generalize \eqref{eq:LB:hyp} to more general sources that are possibly non-hypergraphical. Another interesting open problem is to characterize $\rsfsR$ for PINs with cycles, thereby improving Theorem~\ref{thm:PIN:T} to allow for cycles.

The bound in \eqref{eq:LB:hyp} can be loose for hypergraphical sources. A trivial example is where $V:=[3]$ and
\begin{align*}
	\RZ_1&:=  (\RX_{\rm{a}},\phantom{\RX_{\rm{b}}}\:\:\RX_{\rm{c}})\\
	\RZ_2&:=  (\RX_{\rm{a}},\RX_{\rm{b}},\RX_{\rm{c}})\\
	\RZ_3&:=  \phantom{\RX_{\rm{a}}}\:\:(\RX_{\rm{b}},\RX_{\rm{c}}).
\end{align*}
The numerator of $\alpha(\mcP)$ in \eqref{eq:LB:hyp:alpha} is 0 for any $\mcP$, as the mininum is achieved by the hyperedge $c$ incident on all the nodes. Hence, $\alpha(\mcP)=0$ and so \eqref{eq:LB:hyp} becomes trivial. However, with $B=\{1,3\}$ and $\mcP=\{\{1\},\{3\}\}$, \eqref{eq:LB} gives $r_2\geq \rK-1$, which is non-trivial for $1<\rK\leq 2=\CS$. We also conjecture that \eqref{eq:LB} and \eqref{eq:LB:hyp} are both loose for the example where $V:=[6]$ and 
\begin{align*}
	\RZ_1 & :=(\RX_{\rm{a}},\phantom{\RX_{\rm{b}},\RX_{\rm{c}},}\:\RX_{\rm{d}})\\
	\RZ_2 & :=(\RX_{\rm{a}},\RX_{\rm{b}}\phantom{,\RX_{\rm{c}},\RX_{\rm{d}}})\\
	\RZ_3 & :=(\RX_{\rm{a}},\RX_{\rm{b}},\phantom{\RX_{\rm{c}},}\:\RX_{\rm{d}})\\
	\RZ_4 & :=(\phantom{\RX_{\rm{a}},}\:\RX_{\rm{b}},\RX_{\rm{c}},\RX_{\rm{d}})\\
	\RZ_5 & :=(\phantom{\RX_{\rm{a}},}\:\RX_{\rm{b}},\RX_{\rm{c}}\:\phantom{\RX_{\rm{d}}}\:)\\
	\RZ_6 & :=\phantom{(\RX_{\rm{a}},\RX_{\rm{b}},}\:\RX_{\rm{c}}.
\end{align*}
We conjecture that $(\rK,r_V)\in\rsfsR$ only if
$$
r(V)\geq 1.5\rK,
$$
which is achievable using the idea of secret key agreement by network coding \cite{chan10md}. It can be shown that the best lower bound from \eqref{eq:LB} and \eqref{eq:LB:hyp} is $r(V)\geq \rK$. Hence, we expect that resolving the conjecture in the affirmative potentially leads to new techniques for obtaining better lower bounds on the public discussion rate required for secret key agreement.

\appendices

\makeatletter
\@addtoreset{equation}{section}
\renewcommand{\theequation}{\thesection.\arabic{equation}}
\renewcommand{\theparentequation}{\thesection.\arabic{parentequation}}
\@addtoreset{Theorem}{section}
\renewcommand{\theTheorem}{\thesection.\arabic{Theorem}}
\@addtoreset{Lemma}{section}
\renewcommand{\theLemma}{\thesection.\arabic{Lemma}}
\@addtoreset{Corollary}{section}
\renewcommand{\theCorollary}{\thesection.\arabic{Corollary}}
\@addtoreset{Example}{section}
\renewcommand{\theExample}{\thesection.\arabic{Example}}
\@addtoreset{Remark}{section}
\renewcommand{\theRemark}{\thesection.\arabic{Remark}}
\@addtoreset{Proposition}{section}
\renewcommand{\theProposition}{\thesection.\arabic{Proposition}}
\@addtoreset{Definition}{section}
\renewcommand{\theDefinition}{\thesection.\arabic{Definition}}
\@addtoreset{Subclaim}{Theorem}
\renewcommand{\theSubclaim}{\theLemma.\arabic{Subclaim}}
\makeatother

\section{Proof of Theorem~\ref{thm:LB}}
\label{sec:proof}

To prove Theorem~\ref{thm:LB}, we will first prove the mult-letter version of the bound in terms of $I_{\mcP}$:
\begin{Lemma}
	\label{lem:ML:LB}
	For any $B\subseteq V$ with size $\abs{B}>1$,
	\begin{align}
		\label{eq:ML:LB}
		\kern-.5em H(\RF_{V\setminus B})\kern-.2em-\kern-.2em H(\RF|\tRZ_B)
		\geq (\abs{\mcP}\kern-.2em-\kern-.2em1)\kern-.2em`1[I_{\mcP}(\tRZ_B|\RF)-I_{\mcP}(\tRZ_B)`2]\kern-.5em
	\end{align}
	for any $\mcP\in \Pi'(B)$.
\end{Lemma}
\begin{Proof}
	Consider any $B\subseteq V$ such that $\abs{B}>1$, and $\mcP\in \Pi'(B)$ as stated in the lemma. Define 
	\begin{align}
		\label{eq:a_t}
		a_t:=I_{\mcP}(\tRZ_B|\RF^t_V)-I_{\mcP}(\tRZ_B|\RF^{t-1}_V)\kern1em\text{for } t\in [\ell],
	\end{align}
	where $\RF^0_V:=0$ deterministically for notational convenience. Then, we have the telescoping sum 
	\begin{align*}
		\sum_{t=1}^{\ell}a_t=I_{\mcP}(\tRZ_B|\RF)-I_{\mcP}(\tRZ_B),
	\end{align*} 
	and so it suffices to show that 
	\begin{align}
		(\abs {\mcP}-1) \sum_{t=1}^{\ell}a_t \leq \text {r.h.s. of~\eqref{eq:ML:LB}.}\label{eq:proof:ML:LB:1}
	\end{align}
	By the definition~\eqref{eq:IP} of $I_{\mcP}$,
	\begin{align}
		a_t&=
		\frac{\sum_{C\in\mcP} H(\tRZ_{C}|\RF^{t}_V)-H(\tRZ_B|\RF^{t}_V)}{\abs{\mcP}-1}\notag\\
		&\kern1em -\frac{\sum_{C\in\mcP} H(\tRZ_{C}|\RF^{t-1}_V)-H(\tRZ_B|\RF^{t-1}_V)}{\abs{\mcP}-1} \notag\\ 
		\begin{split}
		(\abs{\mcP}-1)a_t &=\underbrace{I(\tRZ_B\wedge\RF_{Vt}|\RF^{t-1}_V)}_{`(1)}\\[-2em]
		&\kern6em -\overbrace{\sum_{C\in\mcP} I(\tRZ_{C}\wedge\RF_{Vt}|\RF^{t-1}_V)}^{`(2)},
		\end{split}\label{eq:proof:ML:LB:2}
	\end{align}
	where we have grouped the entropy terms in different brackets into the mutual information terms in the last expression by the definition of conditional mutual information. Using standard techniques (cf.\ \cite[Lemma~B.1]{csiszar08}),
	\begin{align*}
	`(2)&\utag{a}=\sum_{C\in\mcP} \sum_{i\in V} I(\tRZ_{C}\wedge\RF_{it}|\tRF_{it})\notag\\
		&\utag{b}\geq\sum_{C\in\mcP} \sum_{i\in C}H(\RF_{it}|\tRF_{it})\notag\\
		&\utag{c}=\sum_{i\in B}\sum_{C\in\mcP:i\in C} H(\RF_{it}|\tRF_{it})\notag\\
		&\utag{d}=\sum_{i\in B} H(\RF_{it}|\tRF_{it})\notag\\
		&\utag{e}\geq\sum_{i\in B} H(\RF_{it}|\RF^{t-1}_V,\RF_{[i-1]\cap B\,t},\RF_{V`/B\, t})\notag\\
		&\utag{f}=H(\RF_{Bt}|\RF^{t-1}_V,\RF_{V`/B\, t}) 
	\end{align*} 
	\begin{compactitem}
	\item where \uref{a} follows from the chain rule and the definition~\eqref{eq:tFit} of $\tRF_{it}$;
		\item \uref{b} is because
	\begin{align*}
		I(\tRZ_{C}\wedge\RF_{it}|\tRF_{it})
		\begin{cases}
			=H(\RF_{it}|\tRF_{it})  & \text {if $i \in C$ by \eqref{eq:Fit},}\\
			\geq 0 & \text{otherwise};
		\end{cases}
	\end{align*}
	\item \uref{c} is obtained by interchanging sums;
	\item \uref{d} is because the summand on r.h.s.\ of \uref{c} is constant w.r.t.\  $C$, and so the inner summation gives a multiplicative factor of $1$.
	\item \uref{e} is obtained by \eqref{eq:tFit} and an additional conditioning on $\RF_{V`/B\, t}$, which does not increase the entropy.
	\item \uref{f} follows from the chain rule.
	\end{compactitem}
Hence,
\begin{align*}
	`(1)-`(2) &\leq `1[H(\RF_{Vt}|\RF_V^{t-1}) -H(\RF_{Vt}|\RF_V^{t-1},\tRZ_B)`2]\\
	&\kern1em - `1[H(\RF_{Vt}|\RF_V^{t-1}) -H(\RF_{V`/B\,t}|\RF_V^{t-1})`2]\\
	&=  \underbrace{H(\RF_{V`/B\,t}|\RF_V^{t-1})}_{\leq b_t:=H(\RF_{V`/B\,t}|\RF_{V`/B}^{t-1})} - \underbrace{H(\RF_{Vt}|\RF_V^{t-1},\tRZ_B)}_{c_t}
\end{align*}
Since $\sum_{t=1}^{\ell} b_t=H(\RF_{V`/B})$ and $\sum_{t=1}^{\ell} c_t=H(\RF_V|\tRZ_B)$ by the chain rule, the above inequality and \eqref{eq:proof:ML:LB:2} gives
\begin{align*}
	(\abs {\mcP}-1)\sum_{t=1}^{\ell} a_t &\leq H(\RF_{V`/B}) - H(\RF_V|\tRZ_B),
\end{align*}
which establishes \eqref{eq:proof:ML:LB:1} as desired.
\end{Proof}

We now single-letterize \eqref{eq:ML:LB} to give the desired lower bound~\eqref{eq:LB} in Theorem~\ref{thm:LB}:
\begin{Proof}[Theorem~\ref{thm:LB}]
	Consider any $B\subseteq V$ with size $\abs{B}>1$ and $\mcP\in\Pi'(B)$ as stated in the theorem. 
	l.h.s.\ of~\eqref{eq:ML:LB} in Lemma~\ref{lem:ML:LB} can be bounded by the total discussion rate as follows:
	\begin{align}
		\kern-.5em H(\RF_{V`/B}) - H(\RF_V|\tRZ_B) &\leq H(\RF_{V`/B}) \leq \sum_{i \in V`/B} \log \abs{F_i} \kern-2em\notag\\
		&\leq n `1[r(V`/B)+`d^{(1)}_n`2]  \label{eq:proof:LB:1}
	\end{align}
	for some $`d^{(1)}_n\to 0$ as $n\to `0$ by \eqref{eq:rate}. Next, we simplify first term on the r.h.s.\ of \eqref{eq:ML:LB} as follows:
	\begin{align}
		\kern-.6em I_{\mcP}(\tRZ_B|\RF)
		&\utag{a}= \frac{\sum_{C\in\mcP}H(\tRZ_C|\RF)-H(\tRZ_B|\RF)}{\abs{\mcP}-1}  \notag\\
		&\utag{b}\geq H(\RK|\RF) + I_{\mcP}(\tRZ_B|\RF,\RK)-n`d^{(2)}_n\notag\\
		&\utag{c}\geq n(\rK-`d^{(2)}_n-`d^{(3)}_n)\label{eq:proof:LB:2}
	\end{align}
\begin{compactitem}
	\item where \uref{a} is by the definition~\ref{eq:IP} of $I_{\mcP}$;
	\item \uref{b} is obtained by applying the inequalities
	\begin{align*}
		H(\tRZ_C|\RF) + n`d^{(2)}_n \tfrac{\abs {\mcP}-1}{\abs {\mcP}}&\geq H(\RK,\tRZ_C|\RF)\\
		&= H(\RK|\RF)+H(\tRZ_C|\RF,\RK)
	\end{align*}
	for some $`d^{(2)}_n\to 0$, by \eqref{eq:recover} and Fano's inequality, and
	\begin{align*}
	H(\tRZ_B|\RF) &\leq H(\RK,\tRZ_B|\RF)\\
	&= H(\RK|\RF)+H(\tRZ_B|\RF,\RK),
	\end{align*}
	and then grouping the entropy terms involving $\tRZ_C$ to form $I_{\mcP}(\tRZ_B|\RF,\RK)$;
	\item \uref{c} is because $I_{\mcP}(\tRZ_B|\RF,\RK)\geq 0$ by the positivity of divergence in \eqref{eq:IP}, and $H(\RK|\RF)\geq n[\rK-`d^{(3)}_n]$ for some $`d^{(3)}_n\to 0$ by \eqref{eq:secrecy}.
\end{compactitem}
	Finally, the last term on the r.h.s.\ of \eqref{eq:ML:LB} can be single-letterized as follows:
	\begin{align}
		I_{\mcP}(\tRZ_B)
		&\utag{d}= \frac{\sum_{C\in\mcP}H(\tRZ_C)-H(\tRZ_B)}{\abs{\mcP}-1}  \notag\\
		&\utag{e}= \frac{\sum_{C\in\mcP} \sum_{i\in C} H(\RU_i)-\sum_{i \in B} H(\RU_i)}{\abs{\mcP}-1}  \notag\\
		&\kern1em + \frac{\sum_{C\in\mcP}nH(\RZ_C)-nH(\RZ_B)}{\abs{\mcP}-1}\notag\\
		&\utag{f}=nI_{\mcP}(\RZ_B)\label{eq:proof:LB:3}
	\end{align}
\begin{compactitem}
	\item where \uref{d} is by the definition~\eqref{eq:IP} of $I_{\mcP}$;
	\item \uref{e} is obtained by the expansion
	\begin{align*}
		H(\tRZ_C) &= H(\RU_C,\RZ_C^n) = \sum_{i \in C} H(\RU_i)+n H(\RZ_C)\\
		H(\tRZ_B) &= H(\RU_B,\RZ_B^n) = \sum_{i \in B} H(\RU_i)+n H(\RZ_B)
	\end{align*}
	by the definition~\eqref{eq:tRZ} of $\tRZ_V$, the independence assumption~\eqref{eq:U} and the fact that $\RZ_V^n$ is i.i.d.\ generated from the source $\RZ_V$; 
	\item \uref{f} is because the expression in the first pair of brackets evaluates to $0$ by exchanging the first two summation, and the expression in the second pair brackets evaluate to $nI_{\mcP}(\RZ_B)$.
\end{compactitem} 
Applying \eqref{eq:proof:LB:1}, \eqref{eq:proof:LB:2} and \eqref{eq:proof:LB:3} to \eqref{eq:ML:LB} and dividing both sides by $n$, we have the desired lower bound~\eqref{eq:LB} in the limit as $n\to `8$.
\end{Proof}

\section{Proof of Theorem~\ref{thm:LB:hyp}}
\label{sec:proof:hyp}

\def\u{1em}
\def\MarkLt{0.4*\u}
\def\MarkSep{0.2*\u}
\tikzset{
	TwoMarks/.style={
		postaction={decorate,
			decoration={
				markings,
				mark=at position #1 with
				{
					\begin{scope}[xslant=0.2]
						\draw[line width=\MarkSep,white,-] (0pt,-\MarkLt) -- (0pt,\MarkLt) ;
						\draw[-] (-0.5*\MarkSep,-\MarkLt) -- (-0.5*\MarkSep,\MarkLt) ;
						\draw[-] (0.5*\MarkSep,-\MarkLt) -- (0.5*\MarkSep,\MarkLt) ;
					\end{scope}
				}
			}
		}
	},
	TwoMarks/.default={0.5},
}

\tikzstyle{point}=[minimum size=0em,inner sep=0, outer sep=0em]
\tikzstyle{htag}=[draw,rectangle,minimum width=.4*\u,minimum height=0,inner sep=0, outer sep=0em]
\tikzstyle{vtag}=[draw,rectangle,minimum height=.4*\u,minimum width=0,inner sep=0, outer sep=0em]
\tikzstyle{aline}=[dashed,gray]
\tikzstyle{bar}=[fill=cyan!10!white!90,draw=cyan]

\begin{figure*}
	\begin{center}
	\subcaptionbox{$`m^*$ in general~\eqref{eq:`m*}.\label{fig:greedy:general}}{
	\def\ux{3em}
	\def\uy{2.5em}
	\tikzstyle{point}=[minimum size=0em,inner sep=0, outer sep=0em]
	\tikzstyle{htag}=[draw,rectangle,minimum width=.4*\u,minimum height=0,inner sep=0, outer sep=0em]
	\tikzstyle{vtag}=[draw,rectangle,minimum height=.4*\u,minimum width=0,inner sep=0, outer sep=0em]
	\tikzstyle{aline}=[dashed,gray]
	\tikzstyle{bar}=[fill=cyan!10!white!90,draw=cyan]
	\begin{tikzpicture}[>=latex]
	\scriptsize
	\node [point] (0) at (0,0) {};
	
	\foreach \j/\p/\w/\wl/\sl in {%
		1/0/6/{$w_{s_1}$}/{$s_1$},%
		3/1.2/4/{$w_{s_j}$}/{$s_{j}$},%
		4/2.2/3/{$w_{s_{j+1}}$}/{$s_{j+1}$},%
		7/3.4/1/{$w_{s_k}$}/{$s_k$}%
	}
	{
		\draw[bar] (\p*\ux,0) -- +(0,\w*\uy) node [point] (\j) {} -- +(1*\ux,\w*\uy)  -- +(1*\ux,0) -- cycle;
		\path (\j)  -- (0|-\j) node (w\j) [htag,label={[label distance=0.1*\u]180:\wl}] {};
		\draw[aline] (\j) -- (w\j.center);
		\path (\j|-0) to node [vtag,label={[label distance=0.4*\u]-90:\sl}] {} +(1*\ux,0);
	};
	
	\foreach \j/\lb in {%
		3/$\kern3.2em S_j:=\Set{s_{j'}\mid  j'\leq j}$%
		,7/$S_k=S$%
	}
	\draw[decorate, decoration={brace, amplitude=3pt,mirror,raise=0.1*\u}] (\j) +(\ux,0)  node (S\j) {}  -- (0|-S\j) node [midway,above,yshift=0.2*\u] {\contour{white}{\lb}}; 
	

	\foreach \j/\lb in {%
		3/$`m^*(S_j):=w_{s_j}-w_{s_{j+1}}$%
	}
	\draw[|<->|] (\j) +(1.1*\ux,0) -- +(1.1*\ux,-\uy) node [midway,right,xshift=0.2*\u] {\contour{white}{\lb}} ; 
	
	\draw[|<->|] (7) +(1.1*\ux,0) -- +(1.1*\ux,-\uy) node [midway,right,xshift=0.2*\u] {$`m^*(S):=w_{s_k}\kern-1em $} ; 
	
	\path (0) to node  {}  (4.8*\ux,0) node (s) [point,label={[label distance=0.1*\u]right:$s\in S$}] {};
	\path (0) -- (0,7*\uy) node (w) [point,label={[label distance=0.1*\u]above:$w_s$}] {};
	
	\draw[-,TwoMarks=0.675] (0)-- (4-|0);
	\draw[->,TwoMarks] (4-|0)-- (w);
	\path (0) -- +(\ux,0) node [point] (bx1) {};
	\draw[-] (0) -- (bx1);
	\draw[-,TwoMarks] (bx1)-- (3|-0);
	\path (0-|4) -- +(\ux,0) node [point] (bx2) {};
	\draw[-,TwoMarks] (bx2)-- (7|-0);
	\draw[-] (3|-0)-- (bx2);
	\draw[-] (7|-0)-- (s);
	
	\end{tikzpicture}
	\vspace{2.2em}
}\hfill
	\subcaptionbox{$`m^*$ applied to the proof of~\eqref{eq:greedy:proof}.\label{fig:greedy:proof}}{
	\def\ux{3em}
	\def\uy{2.5em}
	\begin{tikzpicture}[>=latex]
	\scriptsize
	\node [point] (0) at (0,0) {};
	
	\foreach \j/\p/\w/\wl/\sl in {%
		1/0/6/{$w_{0}=\abs {\mcP}$}/{$s_1=0$},%
		2/1/5/{$w_{e_1}$}/{$s_2=e_1$},%
		3/2.2/4/{$w_{e_j}$}/{$\begin{aligned}&s_{j+1}\\[-.5em] &\kern.5em= e_j\end{aligned}$},%
		4/3.2/3/{$w_{e_{j+1}}$}/{$\begin{aligned}&s_{j+2}\\[-.5em] &\kern.5em= e_{j+1}\end{aligned}$},%
		5/4.4/2/{$w_{e_{\abs{E}}}$}/{$\begin{aligned}&s_{\abs{E}+1}\\[-.5em] &\kern.5em= e_{\abs{E}}\end{aligned}$},%
		6/5.4/1/{}/{$s_{\abs {E}+2}$},%
		7/6.6/1/{$1$}/{$s_{\abs {E}+\abs {V}+1}$}%
	}
	{
		\draw[bar] (\p*\ux,0) -- +(0,\w*\uy) node [point] (\j) {} -- +(1*\ux,\w*\uy)  -- +(1*\ux,0) -- cycle;
		\path (\j)  -- (0|-\j) node (w\j) [htag,label={[label distance=0.1*\u]180:\wl}] {};
		\draw[aline] (\j) -- (w\j.center);
		\path (\j|-0) to node [vtag,label={[label distance=0.4*\u,rotate=-45]0:\sl}] {} +(1*\ux,0);
	};
	
	\foreach \j/\y/\lb in {%
		1/1/$S_1=\Set{0}$%
		,3/3/$S_{j+1}=\Set{0}\cup \Set{e_{j'}\mid j'\leq j}$%
		,5/-1/$S_{\abs{E}+1}=\Set{0}\cup E$%
		,7/-1/$S_{\abs{V}+\abs{E}+1}=S$%
	}
	\draw[decorate, decoration={brace, amplitude=3pt,mirror,raise=0.1*\u}] (\j) +(\ux,0)  node (S\j) {}  -- (0|-S\j) node [midway,above,yshift=0.2*\u,xshift=\y*\u] {\contour{white}{\lb}}; 
	
	\draw[decorate, decoration={brace, amplitude=3pt,mirror,raise=0.4*\u}] (7|-0) +(\ux,0)  node (S7) {}  -- (6|-S7) node [midway,above,yshift=0.6*\u] {$V$};

	\foreach \j/\lb in {%
		1/{$`m^*(S_1)=\abs{\mcP}-w_{e_1}$}%
		,3/$`m^*(S_{j+1})=w_{e_{j}}-w_{e_{j+1}}$%
		,5/$`m^*(S_{\abs{E}+1})=w_{e_{\abs{E}}}-1$%
	}
	\draw[|<->|] (\j) +(1.1*\ux,0) -- +(1.1*\ux,-\uy) node [midway,right,xshift=0.2*\u] {\lb} ; 
	
	\draw[|<->|] (7) +(1.1*\ux,0) -- +(1.1*\ux,-\uy) node [midway,right,xshift=0.2*\u] {$`m^*(S)=1$} ; 
	
	\path (0) to node  {}  (8*\ux,0) node (s) [point,label={[label distance=0.1*\u]right:$s\in S$}] {};
	\path (0) -- (0,7*\uy) node (w) [point,label={[label distance=0.1*\u]above:$w_s$}] {};
	
	\draw[-] (0) --(5-|0);
	\draw[-,TwoMarks] (5-|0)-- (4-|0);
	\draw[-] (4-|0) --(3-|0);
	\draw[-,TwoMarks] (3-|0)-- (2-|0);
	\draw[->] (2-|0)-- (w);
	
	\draw[-] (0) --(2|-0);
	\draw[-,TwoMarks] (2|-0)-- (4|-0);
	\draw[-,TwoMarks] (4|-0)-- (6|-0);
	\draw[-,TwoMarks] (6|-0)-- (S7|-0);
	\draw[-] (S7|-0)-- (s);
	
	\end{tikzpicture}
}
\end{center}
	\caption{Illustration of Edmonds' greedy algorithm in Lemma~\ref{pro:greedy}.}
	\label{fig:greedy}
\end{figure*}

To prove Theorem~\ref{thm:LB:hyp}, we will make use of Edmonds' greedy algorithm in combinatorial optimization~\cite{schrijver02}. A set function $f:2^S\to `R$ with a finite ground set $S$ is said to be \emph{submodular} iff for all $B_1,B_2\subseteq S$,
\begin{align}
	f(B_1)+f(B_2) \geq f(B_1\cap B_2)+f(B_1\cup B_2).\label{eq:submodular}
\end{align}
$f$ is said to be \emph{supermodular} if $-f$ is submodular. If $f$ is both submodular and supermodular, it is said to be \emph{modular}. $f$ is said to be \emph{normalized} if $f(`0)=0$. The entropy function $B\mapsto H(\RZ_B)$~\cite{yeung08}, for instance, is a well-known normalized submodular function~\cite{fujishige78}. Edmonds' greedy algorithm states that:

\begin{Proposition}[\mbox{\cite[Theorem~44.3]{schrijver02}}]
	\label{pro:greedy}
	For any normalized submodular function $f:2^S\to `R$ with a finite ground set $S$, and any non-negative weight vector $w_S:=(w_s\mid s\in S)\in `R_+^S$, 
	consider the linear program
	\begin{subequations}
		\label{eq:sfo}
		\begin{align}
			&\min_{`m} \sum_{B\subseteq S} `m(B)f(B) \label{eq:sfo:obj}
		\end{align}
	such that $`m:2^S\to `R_+$ is a non-negative set function satisfying
		\begin{align}
			&\sum_{B\subseteq S \colon s\in S }`m(B)=w_s, \kern1em\forall s\in S.  \label{eq:sfo:cons}
		\end{align}
	\end{subequations}
	Then, the optimal solution $`m^*$ to the above problem is given as follows:
	\begin{enumerate}
		\item\label{enum:greedy:1} Enumerate $S$ as $\Set{s_1,...,s_k}$ (with $k:=\abs {S}$) such that 
		$$w_{s_1}\geq \dots \geq w_{s_k}.$$
		\item\label{enum:greedy:2} With $S_j:=\Set{s_{j'}\mid 1\leq j'\leq j}$ for $1\leq j\leq k$, set
		\begin{subequations}
			\label{eq:`m*}
		\begin{align}
			`m^*(S_j)&:=w_{s_j}-w_{s_{j+1}}\kern1em \text {for $1\leq j< k$}\\
			`m^*(S_k)&:=`m^*(S)=w_{s_k}
		\end{align}
		\end{subequations}
		and $`m^*(B)=0$ otherwise, i.e., if $B\neq S_j$ for $1\leq j\leq k$.
	\end{enumerate}
	It follows that, if $f$ is modular, the summation in \eqref{eq:sfo:obj} is constant for all feasible $`m$ satisfying \eqref{eq:sfo:cons}.\footnote{This is because $-f$ is submodular and so the same $`m^*$ defined in \eqref{eq:`m*} both minimizes and maximizes the sum in \eqref{eq:sfo:obj}, the value of which must therefore be a constant.}
\end{Proposition}
The algorithm is illustrated in \figref{fig:greedy:general}, which is a plot of $w_s$ against $s\in S$. In particular, the horizontal axis enumerates the elements $S$ in a descending order of their weights $w$ as desired by the greedy algorithm in Step~\ref{enum:greedy:1}. The set of first $j$ elements form the set $S_j$, and the $`m^*(S_j)$ is the drop in height from the $j$-th bar to the $(j+1)$-th bar, with the exception that $`m^*(S_k)$ (or equivalently $`m^*(S)$) is the height of the last bar. 

The proof is by a lamination procedure that can turn any $`m$ to $`m^*$ gradually without increasing the sum in \eqref{eq:sfo:obj} or violating \eqref{eq:sfo:cons}: 
\begin{lbox}
	\noindent\underline{Lamination:}
	For every $B_1,B_2\in \op{supp}(`m)$ such that $B_1$ crosses $B_2$ in the sense that 
	\begin{align*}
		\Set{B_1,B_2}\neq \Set{B_1\cap B_2,B_1\cup B_2},
	\end{align*}
	reduce $`m(B_1)$ and $`m(B_2)$ by $`d$ and increase $`m(B_1\cap B_2)$ and $`m(B_1\cup B_2)$ by $`d$, where
	\begin{align*}
		`d:=\min\Set{`m(B_1),`m(B_2)}\geq 0,
	\end{align*}
	where the non-negativity is by the assumption that $`m$ is non-negative.
	Doing so reduces $\sum_{B\subseteq S} `m(S) f(S)$ by
	\begin{align*}
		`d[f(B_1)+f(B_2)-f(B_1\cap B_2)-f(B_1\cup B_2)]\geq 0,
	\end{align*}
	where the non-negativity is by the submodularity~\eqref{eq:submodular} of $f$.
\end{lbox} 
The procedure turns the support of $`m$ to that of $`m^*$, namely $\Set{S_j\mid 1\leq j\leq k}$, which forms a laminar family (or more specifically, a chain).

\begin{Proof}[Theorem~\ref{thm:LB:hyp}]
	For any $\mcP\in \Pi'(V)$, by \eqref{eq:recover} and Fano's inequality, 
	\begin{align}
	\kern-.5em 	n`d_n &\geq \sum_{C\in\mcP}  H(\RK|\tRZ_{C},\RF) \notag\\
		&= \underbrace{\sum_{C\in\mcP} \kern-.2em H(\tRZ_{C},\RF,\RK)}_{`(1)}
		-\underbrace{\sum_{C\in\mcP} \kern-.2em H(\tRZ_{C})}_{`(2)}
		-\underbrace{\sum_{C\in\mcP} \kern-.2em H(\RF|\tRZ_{C})}_{`(3)} \label{eq:LB:hyp:123}
	\end{align}
	for some $`d_n\to 0$ as $n\to `8$, where the last equality is by the chain rule expansion. We will bound $`(1)$, $`(2)$ and $`(3)$ to obtained the desired lower bound~\eqref{eq:LB:hyp}.
	
	$`(3)$ can be bounded by the usual technique (cf.\ \cite[Lemma~B.1]{csiszar08}):
	\begin{align}
	`(3) &\utag{a}=\sum_{C\in\mcP} \sum_{t=1}^{\ell}\sum_{i\in V}H(\RF_{it}|\tRF_{it},\tRZ_{C}) \notag\\
	&\utag{b}\leq \sum_{C\in\mcP}\sum_{t=1}^{\ell}\sum_{i \in V`/C}H(\RF_{it}|\tRF_{it})\notag\\ 
	&\utag{c}=\sum_{t=1}^{\ell} \sum_{i\in V}\sum_{C\in\mcP\colon i\not\in C} H(\RF_{it}|\tRF_{it})\notag\\ 
	&\utag{d}=(\abs{\mcP}-1)\sum_{t=1}^{\ell}\sum_{i\in V} H(\RF_{it}|\tRF_{it}) \notag\\ 
	&\utag{e}=(\abs{\mcP}-1)H(\RF)\label{eq:se_expansion2}
\end{align}
\begin{compactitem}
	\item where \uref{a} follows from the chain rule expansion on $\RF$~\eqref{eq:discussion};
	\item \uref{b} is because
	\begin{align*}
		H(\RF_{it}|\tRF_{it},\tRZ_{C})
		\begin{cases}
			=0  & \text {if $i \in C$ by \eqref{eq:Fit},}\\
			\leq H(\RF_{it}|\tRF_{it}) & \text{otherwise};
		\end{cases}
	\end{align*}
	\item \uref{c} is obtained by interchanging sums;
	\item  \uref{d} is because the summand on r.h.s.\ of \uref{c} is constant w.r.t.\  $C$, and so the inner summation gives a multiplicative factor of $\abs {\mcP}-1$.
	\item \uref{e} follows again from the chain rule expansion on $\RF$~\eqref{eq:discussion}.
\end{compactitem}
 
 	Next, we will bound $`(1)$ and $`(2)$ using Edmonds' greedy algorithm in Proposition~\ref{pro:greedy}. 	For notational simplicity, define 
 	\begin{align*}
 		E_i&:=\{e\mid i\in `x(e)\} &\kern1em &\text {for }i\in V\\
 		E_C&:=\bigcup_{i\in C} E_i && \text {for }C\subseteq V,
 	\end{align*}
 	which denote the collection of edges incident on node~$i\in V$ and nodes in $C\subseteq V$ respectively.
 	Let $S=\{0\}\cup V\cup E$, where we assume $0\not\in V\cup E$ without loss of generality. Define $\RY_{S}$ with
 	\begin{subequations}
 		\label{eq:Y}
	\begin{alignat}{2}
		\RY_0&=(\RF,\RK) \\
		\RY_i&=\RU_i &\kern1em& \text{for }i\in V \label{eq:Yi}\\ 
		\RY_e&=\RX_e^n && \text{for }e\in E. \label{eq:Ye}
	\end{alignat}
	\end{subequations}
	Note that $\tRZ_C=(\RU_C,\RZ^n_{E_C})=(\RU_C,\RX^n_{E_C})$, where the first equality is by~\eqref{eq:tRZ}, and the second equality is by \eqref{eq:Xe}. Hence, we can rewrite $`(1)$ as the sum $\sum_{B\subseteq S}`m(B)f(B)$ in \eqref{eq:sfo:obj} with
	\begin{align*}
		f(B)&:=H(\RY_B) \kern1em \text{for}\kern1em B\subseteq S.\\
		`m(B)&:=
		\begin{cases} 
			1, &B=\Set{0}\cup C\cup E_C, C\in \mcP \\
			0, & \text{otherwise}.
		\end{cases}
	\end{align*} 
	Then, $f$ is normalized and submodular as it is an entropy function of $\RY_S$~\cite{fujishige78}, and \eqref{eq:sfo:cons} holds with the non-negative weights defined as
	\begin{subequations}
		\label{eq:greedy:proof:w}
	\begin{alignat}{2}
		w_0&:=\sum_{B\subseteq S:0\in S}`m(B) \notag\\
		&=\sum_{C\in \mcP} `m(\Set{0}\cup C\cup E_C)=\abs{\mcP},  \\
		w_i&:=\sum_{B\subseteq S:i\in S}`m(B)&\kern1em& \text{for $i\in V$}\notag\\
		&=\sum_{C\in\mcP:i\in C} `m(\Set{0}\cup C\cup E_C) =1 \\
		w_e&:=\sum_{B\subseteq S:e\in S}`m(B)&&\text{for $e\in E$}\notag\\
		&=\sum_{C\in\mcP:e\in E_C} `m(\Set{0}\cup C\cup E_C)\notag \\ 
		&=|\Set{C\in \mcP\mid C\cap `x(e)\neq `0}|.\label{eq:w_e}
	\end{alignat}
	\end{subequations}
	As an example, for the triangle PIN $\RZ_{\Set{1,2,3}}$ defined in \eqref{eq:triangle} and illustrated in \figref{fig:triangle}, and the partition $\mcP:=\Set {\Set {1},\Set {2},\Set {3}}$ into singletons, 
	\begin{align*}
		w_0&=\abs {\mcP}=3\\
		w_1&=w_2=w_3=1\\
		w_a&=w_b=w_c=2,
	\end{align*}
	as $w_e$ in \eqref{eq:w_e} reduces to the number of incident nodes of edge $e$ for singleton partition.
	 
	It follows that
	\begin{align*}
		w_0 =\abs {\mcP}\geq w_e \geq 1= w_i  \kern1em \forall e\in E, i\in V.
	\end{align*}
	Enumerate $E$ as $\Set{e_1,\dots,e_{\abs{E}}}$ such that
	\begin{align}
		w_{e_1} \geq w_{e_2} \geq \dots \geq w_{e_{\abs {E}}}.\label{eq:greedy:proof:worder}
	\end{align}
	Then, the desired ordering in Step~\ref{enum:greedy:1} of the greedy algorithm in Proposition~\ref{pro:greedy} satisfies
	\begin{subequations}
		\label{eq:greedy:proof:s}
		\begin{align}
			s_1&=\Set{0}\\
			\Set{s_2,\dots s_{\abs{E}+1}}&=\Set{e_1,\dots,e_{\abs{E}}}\\
			\Set{s_{\abs{E}+2},\dots s_{\abs{E}+\abs{V}+1}}&=V
		\end{align}
	\end{subequations}
	and so $`m^*$ defined in~\eqref{eq:`m*} can be evaluated as shown in \figref{fig:greedy:proof}, with possibly non-zero values at
	\begin{alignat*}{2}
		S_1&=\Set{s_1}=\Set{0}\\
		S_{j+1} &= \Set {0}\cup \Set{e_{j'}\mid 1\leq j'\leq j} &\kern1em & \text {for }1\leq j\leq \abs {E}\\
		S_k &= S = \Set{0}\cup E\cup V.
	\end{alignat*}
	By Proposition~\ref{pro:greedy}, we can lower bound $`(1)$ with $\sum_{B\subseteq S} `m^*(B)f(B)$ , which simplifies to
	\begin{align}
		\label{eq:greedy:proof}
		\begin{split}
	`(1)&\geq \overbrace{(\abs {\mcP}-w_{e_1})}^{`m^*(S_1)} H(\overbrace{\RF,\RK}^{\RY_{S_1}})\\
	&\kern1em+\sum_{j=1}^{\abs{E}-1}
	\overbrace{`1(w_{e_j}-w_{e_{j+1}}`2)}^{`m^*(S_{j+1})} H(\overbrace{\RF,\RK,\RX_{\Set{e_{j'}\mid 1\leq j'\leq j}}^n}^{\RY_{S_{j+1}}})\\
	&\kern1em+
	\overbrace{(w_{e_{\abs{E}}}-1)}^{`m^*(S_{\abs {E}+1})}
	H(\overbrace{\RF,\RK,\RX_E^n}^{\RY_{S_{\abs {E}+1}}})+ H(\overbrace{\RF,\RK,\RX_{E}^n,\RU_{V}}^{\RY_S}).
	\end{split}
	\end{align}
	Using the triangle PIN and singleton partition again as an example, we have
	\begin{align*}
		`m^*(\Set {0})=`m^*(\Set {0,\rm{a},\rm{b},\rm{c}}) = `m^*(\Set {0,\rm{a},\rm{b},\rm{c},1,2,3})=1
	\end{align*}
	the above inequality evaluates to
	\begin{align*}
		&H(\tRZ_1,\RF,\RK)+H(\tRZ_2,\RF,\RK)+H(\tRZ_3,\RF,\RK)\\
		&\kern.2em \geq H(\RF,\RK)+H(\RX_{\Set{\rm{a},\rm{b},\rm{c}}},\RF,\RK)
		+H(\RU_{\Set{1,2,3}},\RX_{\Set{\rm{a},\rm{b},\rm{c}}},\RF,\RK).
	\end{align*}
	
	We can follow a similar argument to bound $`(2)$. Note that the entropy in $`(2)$ is the same as that in $`(1)$ except it does not have $(\RF,\RK)$, and so
	we can eliminate $\RY_0$ from the above argument to obtain
	\begin{align*}
		`(2)&= \sum_{j=1}^{\abs{E}-1}`1(w_{e_j}-w_{e_{j+1}}`2)H(\RX_{\Set{e_{j'}\mid 1\leq j'\leq j}}^n)\\
		&\kern1em+(w_{e_{\abs{E}}}-1)H(\RX_E^n)+ H(\RX_{E}^n,\RU_{V}),
	\end{align*}
	which is identical to \eqref{eq:greedy:proof} except that $(\RF,\RK)$ is removed from every entropy term.
	We also have equality here because $f$ is modularover $V\cup E$ due to the fact that $\RY_s$ for $s\in V\cup E$~defined in \eqref{eq:Yi} and \eqref{eq:Ye} are mutually independent because of \eqref{eq:U} and the independence of the edge variables. It follows that
	\begin{align*}
		`(1)-`(2) &\geq (\abs {\mcP}-w_{e_1})H(\RF,\RK) \\
		&\utag{f}= (\abs {\mcP}-1)`1[1-`a(\mcP)`2] H(\RF,\RK)\\
		&\utag{g}= (\abs {\mcP}-1) `1[1-`a(\mcP)`2] `1[H(\RF)+H(\RK)-n`d'_n`2]
	\end{align*}
	for some $`d'_n\to 0$ as $n\to `8$, where
	\begin{compactitem}
		\item	\uref{f} is because by \eqref{eq:greedy:proof:worder} and \eqref{eq:w_e},
		\begin{align*}
			w_{e_1} := \max_{e\in E} w_e &= \max_{e\in E} \abs {\Set {C\in \mcP\mid C\cap `x(e)\neq `0}}\\
			&= (\abs {\mcP}-1)`a(\mcP)+1 \kern1em  \text {by \eqref{eq:LB:hyp:alpha}.}\\
			\abs {\mcP}-w_{e_1}&= (\abs {\mcP}-1)`1[1-`a(\mcP)`2] 
		\end{align*}
		\item \uref{g} is by the secrecy constraint~\eqref{eq:secrecy}.
	\end{compactitem}
	Applying the above inequality and \eqref{eq:se_expansion2} to \eqref{eq:LB:hyp:123} and simplifying, we have
	\begin{align*}
		`a(\mcP)\frac{H(\RF)}n 
		\geq `1[1-`a(\mcP)`2] `1[\frac{H(\RK)}n -`d'_n `2]- \frac{`d_n}{\abs {\mcP}-1},
	\end{align*}
	which implies \eqref{eq:LB:hyp:R} by \eqref{eq:rate} in the limit as $n\to `8$.
\end{Proof}

\section*{Acknowledgment} 
%
%
The authors would like to thank Dr.\ S.\ W. Ho for inspiring discussion and helpful comments.

\bibliographystyle{IEEEtran}
\bibliography{IEEEabrv,ref}

\end{document}